\pdfoutput=1
\documentclass[10pt,letterpaper]{article}
\usepackage{opex3}
\usepackage[export]{adjustbox}
\usepackage{graphicx}
\usepackage{amssymb}
\begin{document}

%%%%%%%%%%%%%%%%%% title page information %%%%%%%%%%%%%%%%%%
\title{Anisotropic model for the fabrication of Annealed and Reverse Proton Exchanged waveguides in congruent Lithium Niobate}

\author{Francesco Lenzini, Sachin Kasture, Ben Haylock, \\ and Mirko Lobino$^{*}$}

\address{Centre for Quantum Dynamics, Griffith University, Brisbane, QLD, Australia\\
and\\
Queensland Micro- and Nanotechnology Centre, Griffith University, Brisbane, QLD, Australia}

\email{$^*$m.lobino@griffith.edu.au} %% email address is required



\begin{abstract}
An anisotropic model for the fabrication of annealed and reverse proton exchange waveguides in lithium niobate is presented. We characterized the anisotropic diffusion properties of proton exchange, annealing and reverse proton exchange in Z-cut and X-cut substrates using planar waveguides. Using this model we fabricated high quality channel waveguides with propagation losses as low as 0.086 dB/cm and a coupling efficiency with optical fiber of 90\% at 1550 nm. The splitting ratio of a set of directional couplers is predicted with an accuracy of $\pm$ 0.06.
\end{abstract}

% REPLACE WITH CORRECT OCIS CODES FOR YOUR ARTICLE, MINIMUM OF TWO; Avoid using the OCIS codes for “General” or “General science” whenever possible.

%%%%%%%%%%%%%%%%%%%%%%% References %%%%%%%%%%%%%%%%%%%%%%%%%
\bibliography{OptExpLenz}
%

%%%%%%%%%%%%%%%%%%%%%%%%%%  body  %%%%%%%%%%%%%%%%%%%%%%%%%%
\section{Introduction}

Lithium niobate (LN) is one of the most widely used materials for the fabrication of integrated optical devices because of its high nonlinearity and large electro-optic coefficient for fast switching \cite{Wooten00}. High speed modulators in LN have been standard components of optical telecommunication networks for decades \cite{Thomsen} and more recently LN integrated optical devices have been used for the optical simulation of solid state systems \cite{longhi06,longhi06b} and for quantum optics experiments including the generation \cite{tanzi} and manipulation \cite{Bonneau12} of single photons and the demonstration of a compact quantum-key-distribution system \cite{Zhang2014}.

LN waveguides are fabricated mainly by two techniques: titanium indiffusion (Ti-indiffusion) \cite{LN_Ti} and proton exchange (either annealed (APE) or reverse (RPE) proton exchange) \cite{Korkishko98}. Ti-indiffused waveguides are fabricated through the deposition of a few nanometer thick Ti layer on top of the material followed by annealing at T$\sim$1000 $^\circ$C in a furnace. This is the standard technique used for commercial modulators and Ti indiffusion has been well characterised and reliably modeled \cite{Ti_model}. For annealed (APE) and reverse (RPE) proton exchange waveguides, the core is fabricated by replacing lithium ions (Li$^+$) with hydrogen ions (H$^+$) by dipping the sample in a hot acid bath: this substitution increases the extraordinary refractive index and decreases the ordinary one \cite{kork2}. In APE waveguides a subsequent annealing step in air is performed to reduce the H$^+$ concentration and improve the optical properties of the waveguide. After proton exchange and annealing a third step of RPE is used to bury the waveguide under the crystal surface and increase the circular symmetry of the optical mode. The main difference between the two techniques is that Ti-indiffused devices guide both polarizations while APE and RPE waveguides guide only light polarized along the optical axis of the crystal.

APE and RPE waveguides have excellent optical properties and they have been used for a broad range of applications in classical and quantum optics including the demonstration of the highest conversion efficiency for second harmonic generation at 1550 nm reported to date \cite{para} and the brightest single photon source based on parametric down conversion ever reported \cite{tanzi}. Nonlinear diffusion models have been proposed for APE \cite{bortz} and RPE \cite{roti1,roti2} processes. The main limitations of these models are the assumptions that the diffusion of H$^+$ ions in the crystal is isotropic even if the crystal is not and that the refractive index changes linearly  with H$^+$ concentration across all the different crystallographic phases \cite{Korkishko98}.

In this paper we report a model of the anisotropic diffusion processes involved in the fabrication of APE and RPE waveguides by studying the H$^+$ diffusion on Z-cut and X-cut substrates. Using this model we derive a relation between proton concentration and refractive index changes as a function of the wavelength when this change is in the $\alpha$-phase ($\Delta n_e <$ 0.025 at 633 nm). This is the phase where the propagation losses of the waveguides are minimized and where the relation of the H$^+$ concentration with the ordinary and extraordinary refractive index changes, $\Delta n_{e,o}$, is linear. We verify the validity of our model with the design and fabrication of channel waveguides and directional couplers for which the anisotropy in the diffusion is particularly critical.

The paper is divided into three more sections: the model of the APE and RPE fabrication steps are described in section 2 for X-cut and Z-cut substrates, in section 3 we compare the prediction of model with the experimental results through the design and fabrication of channel waveguides and directional couplers and section 4 is the conclusion.
\section{Modelling of the refractive index profile evolution in APE and RPE planar waveguides on X-cut and Z-cut substrates}
Lithium niobate is a uniaxial crystal whose optical axis is commonly referred to as the Z axis. In order to characterize the diffusion parameters of APE and RPE processes parallel and orthogonal to this direction we fabricated several planar waveguides on X-cut and Z-cut substrates. The effective refractive indices of the guided modes $n_{eff}$ were measured using the prism coupling technique \cite{TIEN70} with a precision of $\pm$0.0001 set by the $\pm$0.005$^\circ$ angular resolution of our set-up. The optimal diffusion parameters were estimated by minimizing the root-mean-square (rms) error between the measured $n_{eff}$ and those calculated using the nonlinear diffusion model for H$^+$ in LN. For the extraordinary and ordinary refractive indices of bulk LN we used dispersion curves given in \cite{congln} at a temperature $T$=22 $^{\circ}$C.

Only the the extraordinary refractive index $n_e$ increases during proton exchange while the ordinary index $n_o$ decreases. For this reason APE and RPE waveguides guide only one polarization which is TM mode for a Z-cut substrate and TE for X-cut. Nevertheless the ordinary refractive index change $\Delta n_o=-\Delta n_e/3$ was taken into account for the calculation of the modes in Z-cut waveguides \cite{kork2}.

\subsection{Proton exchange}
\begin{figure}[t]
\includegraphics[width = \textwidth] {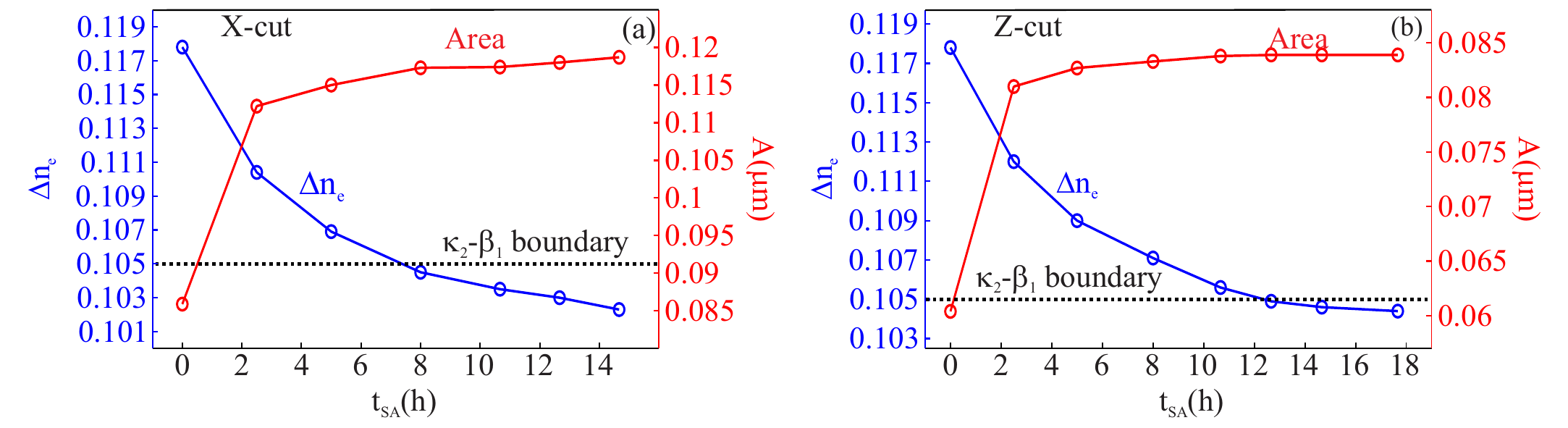}
\caption{Evolution of the refractive index change $\Delta n _e$ and area $A=\Delta n_e \times d_e$ as a function of the SA time for 3 h proton exchanged samples. Circles are experimental data points for (a) X-cut and (b) Z-cut samples. Solid line is plotted as guide to the eye.}
\label{fig_SA}
\end{figure}

Proton exchange (PE) was performed by dipping the LN sample in a pure benzoic acid bath heated to 168.5 $^{\circ}$C. Temperature uniformity is critical for the homogeneity of the fabrication of multiple devices on a single substrate. Our PE reactor is a custom made oil-based heating mantle with internal stirring to improve uniformity.

After PE the samples have a superficial layer at higher refractive index loaded with H$^+$ ions. The guiding layer is formed by different crystallographic phases, with a portion of the protons occupying interstitial sites not contributing to an increase in the refractive index \cite{kork2,roti2,ZavadaAPL}. A soft annealing (SA) step in air at T=210 $^\circ$C follows the PE causing the movement of the interstitial H$^+$ ions to active substitutional sites with a consequent increase in the area of the refractive index profile. The area $A=\Delta n_e \times d_e$ is calculated assuming a step-like refractive index profile of depth $d_e$ and index change $\Delta n_e$. Figure \ref{fig_SA} shows the evolution of the areas $A$ and of the refractive index changes, $\Delta n_e$, with SA time for X-cut and Z-cut planar waveguides that were proton exchanged for 3 h. After each SA step the $n_{eff}$ of the modes were measured by prism coupling and the values of $d_e$ and $\Delta n_e$ was retrieved using the inverse-WKB method. For both samples there is a fast increase of $A$ in the first hours of SA until it saturates to a constant value: this indicates that all interstitial H$^+$ have moved to active substitutional sites. For the Z-cut sample the index change starts to decrease very slowly when $\Delta n_e$ reaches the boundary between the $\kappa_2$ and $\beta_1$ phase corresponding to values below $0.105$ at $\lambda$=635 nm \cite{kork1}. While for Z-cut samples soft annealing acts as a self-stopping process \cite{roti1,roti2}, for X-cut the diffusion proceeds much faster and the values of $\Delta n_e$ shown by the last three data points of Fig. \ref{fig_SA}(a) clearly indicates that the crystal is entering in the $\kappa_2$ phase, where a different relation between proton concentration and $\Delta n_e$ has to be employed.

In our fabrication we stop the SA when the refractive index change, $\Delta n _e$, reaches the boundary between $\kappa_2$ and $\beta_1$ at $0.105$.
Assuming a linear diffusion model for PE, the SA time $t_{SA}$ and the PE time $t_{PE}$ are proportional and given by \cite{roti1,roti2}:
\begin{equation}
t_{SA,X}=2.6  \ \ t_{PE} \ \ \ \ \ \ \mbox{for X-cut samples,}
\label{tsax}
\end{equation}
\begin{equation}
t_{SA,Z}=4  \ \ t_{PE} \ \ \ \ \ \ \mbox{for Z-cut samples,}
\label{tsaz}
\end{equation}

Equations (\ref{tsax}) and (\ref{tsaz}) are valid for a large interval of PE times, ranging from $1 \ \mbox{h}$ to around $10 \ \mbox{h}$. For Z-cut samples, when PE time was larger than 10 h the temperature for SA was increased to a maximum of 230 $^\circ$C and the new SA time calculated as:

\begin{equation}
t_{SA}(T_2)=t_{SA}(T_1) \exp \left( \frac{E_a}{k_b T_2} - \frac{E_a}{k_b T_1}  \right) \ ,
\label{tsaz2}
\end{equation}
where $T_2$ is the new temperature of choice for SA, $T_1=210 \  ^{\circ} \mbox{C}$ is the temperature used for deriving relations (\ref{tsax})-(\ref{tsaz}), $k_b$ is the Boltzmann constant, and $E_a \simeq 1 \ \mbox{eV}$ is the activation energy in the $\beta_1$ phase \cite{roti2}.

\begin{figure}[t]
\centering
\includegraphics[width = 0.8\textwidth] {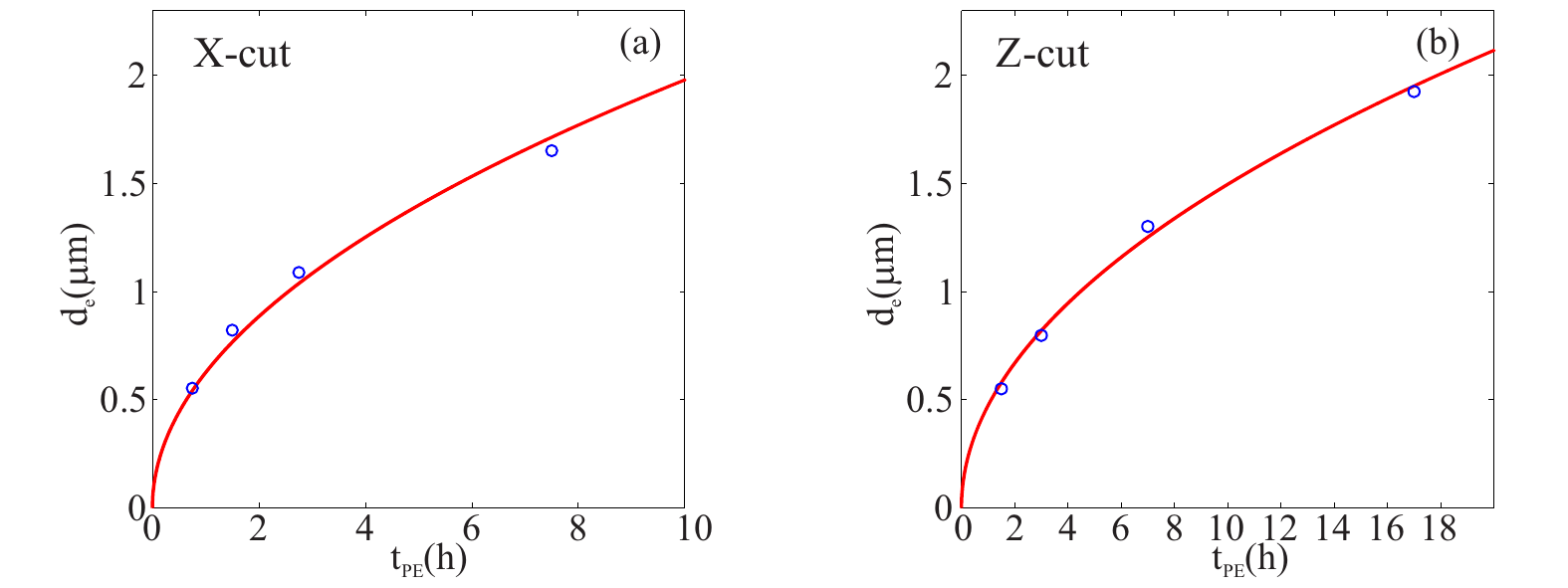}
\caption{Proton exchange depth $d_e$ as a function of time $t_{PE}$ for (a) X-cut and (b) Z-cut sample. Circles are data points fitted with the linear diffusion law $d_e=2\sqrt{D_{PE,X/Z}t_{PE}}$ shown as solid line. The depths are measured after a SA time given by Eqs. (\ref{tsax}) and (\ref{tsaz}).}
\label{fig_PE}
\end{figure}

Relations (\ref{tsax})-(\ref{tsaz2}) have been used to estimate the evolution of the waveguide depth with PE time with $d_e$ ranging from 0.5 $\mu$m to 2 $\mu$m. Figure \ref{fig_PE} shows the measured evolution of $d_e$ with PE time for X-cut and Z-cut waveguides fitted with the linear diffusion law $d_e=2\sqrt{D_{PE,X/Z}t_{PE}}$ \cite{kork1}. From our measurement we found that the diffusion coefficients for the two different substrates are:
\begin{equation}
D_{PE,X} = 0.098 \ \mbox{\textmu m} ^2 \mbox{h}^{-1}\ \ \ \ \ \ \mbox{for X-cut samples,}
\label{dpex}
\end{equation}
\begin{equation}
D_{PE,Z} = 0.056 \ \mbox{\textmu m} ^2 \mbox{h}^{-1}\ \ \ \ \ \ \mbox{for Z-cut samples.}
\label{dpez}
\end{equation}
%Note that these values have to be considered only indicative, as for $d_e > 1 \ \mbox{\textmu m}$, depending on the freshness of the benzoic acid used, fluctuations as large as $\pm 100 \ \mbox{nm}$ can be expected. The SA depth can be however properly monitored during a whole wafer fabrication process, and the following annealing and RPE time chosen to match at best the design of choice.
%
%
\subsection{Annealing}
Proton exchanged waveguides have high propagation losses and a second order susceptibility almost totally suppressed \cite{kork1}. Annealing in dry atmosphere at a temperature of 328 $^\circ$C is performed to decrease the local H$^+$ concentration, $C$, and improve the optical properties of the waveguide in terms of propagation losses, second order nonlinearity and coupling with optical fiber.

The step-like refractive index profile obtained after PE and SA is the initial condition for modelling the annealing diffusion. The proton concentration is set to $C(y)$=1 for $0<y<d_{e}$ and  to 0 for $y>d_{e}$ with $y$ indicating the direction orthogonal to the air-LN interface. The kinetics of H$^{+}$ ions in X-cut and Z-cut planar waveguides are modelled by the one-dimensional nonlinear diffusion equation

\begin{equation}
\frac{\partial C}{\partial t}= \frac{\partial}{\partial y} \left( D_{a,X/Z} \left( C \right) \frac{\partial C}{\partial y} \right),
\label{diffeq}
\end{equation}
where the dependence of the diffusion coefficient $D_{a,X/Z}$ on $C$ is given by \cite{roti1,roti2}:
\begin{equation}
D_{a,X/Z}(C)=D_{0,X/Z} \left( \alpha_{X,Z} + \frac{1-\alpha_{X,Z}}{\beta_{X,Z} C +\gamma_{X,Z}}  \right).
\label{diffcoef}
\end{equation}
\begin{table}[b!]
\centering\caption{Parameters of the diffusion coefficients and Sellmeier curves for X-cut and Z-cut LN substrates.}
\begin{tabular}{c c|c c|c c|c c}
%\hline
\multicolumn{4}{c|}{X-cut}    &  \multicolumn{4}{|c}{Z-cut} \\ \hline
$D_{0}$ ($\mu$m$^2$/h)  & 0.334   & A & 5.063e-3 & $D_{0}$ ($\mu$m$^2$/h) & 0.414  & A & 4.646e-3 \\
$\alpha$ & 0.116   & B & 1.294e-3 & $\alpha$ & 0.134  & B & 9.632e-4 \\
$\beta$  & 30.7    & C($\mu$m) & 0.217   & $\beta$  & 34.5   & C($\mu$m) & 0.272 \\
$\gamma$ & 0.00711 &   &           & $\gamma$ & 0.0497 &   & \\
\end{tabular}
\label{table_diff}
\end{table}
For $\alpha=0$, the rational form of $D_{a,X/Z}(C)$ is the same that would be obtained by a simple inter-diffusion model for $\mbox{H}^{+}$ and $\mbox{Li}^{+}$ ions after the requirement of an electroneutrality condition \cite{vohra}. The additional term $\alpha$ acts as an empirical correcting factor taking into account the different values of the self diffusion coefficients of the two ion species in each phase encountered during  annealing and RPE in a multiphase crystal. When the waveguide is in the $\alpha$-phase the refractive index change is proportional to $C$ and given by $\Delta n_e(\lambda)=\delta_{X/Z}(\lambda) C$. We independently characterize the coefficient $\delta_{X/Z}(\lambda)$ for the two substrates because the stresses experienced by the crystal during the high temperature diffusion processes are different for Z-cut and X-cut waveguides and this may affect the final value of $\Delta n_e$ \cite{kork2}.
\begin{figure}[t]
\includegraphics[width = \textwidth] {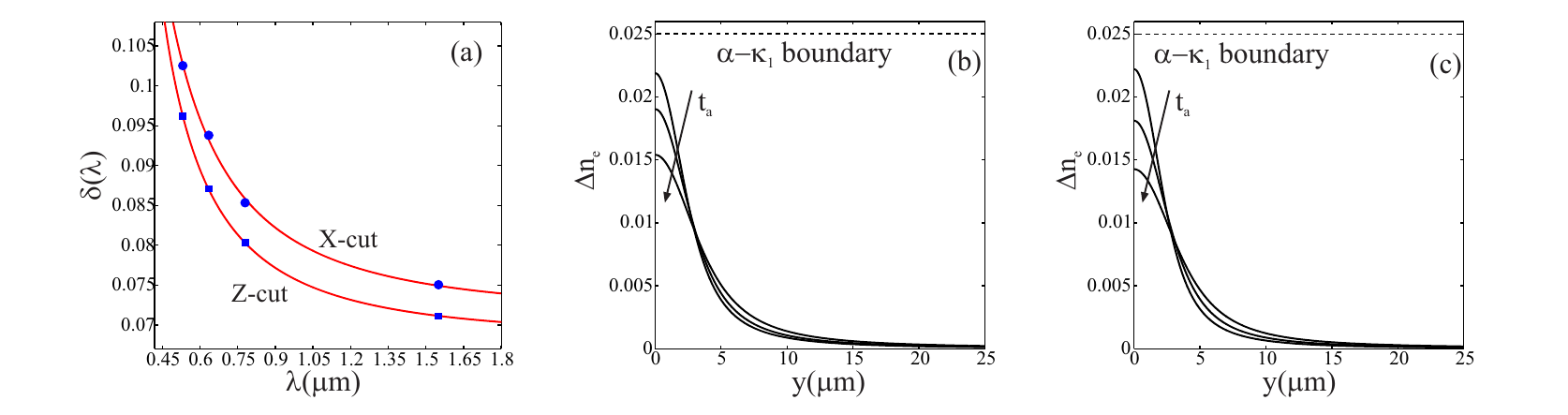}
\caption{(a) Sellmeier fitting (solid lines) and experimental data of the wavelength dependence of the refractive index change coefficient $\delta$  for ($\bullet$) X-cut and ({\tiny $\blacksquare$}) Z-cut. (b) Simulated evolution of the refractive index change $\Delta n_e$ during annealing for an X-cut planar waveguide calculated from Eq. (\ref{diffeq}). The waveguide had a PE depth  $d_e$ = 0.822 $\mu$m and was annealed for $t_a$ = 36 h, 44 h and 59 h. (c) Same as (b) but for Z-cut with $d_e$ = 0.798 $\mu$m and $t_a$ = 25 h, 34 h and 48 h.}
\label{fig_Ann}
\end{figure}
Equation (\ref{diffeq}) is integrated using a semi-implicit finite difference algorithm \cite{Weickert} with $0<y<y_{max}$ and $0<t<t_a$. The boundary conditions are $\frac{\partial C}{\partial y}=0$ at $y=0$, meaning that there is no H$^+$ flux at the LN-air interface, and for $y=y_{max}$ we used a transparent boundary condition in order to simulate an infinitely thick substrate. This is necessary in order to avoid the use of large integration window since for large $t_a$ the refractive index profiles have a long tail that extends for tens of microns.

We fabricated several planar waveguides on the two different substrates with proton exchange depth $d_e$ ranging from 0.5 $\mu$m to $\sim$2 $\mu$m and monitored their evolution during annealing by measuring the $n_{eff}$ of the modes. The values of the parameters $D_{0,X/Z},\alpha_{X/Z},\beta_{X/Z},\gamma_{X/Z}$ and $\delta_{X/Z}$ were determined by minimizing the root-mean-square error between the measured effective indices, $n_{eff}^{meas}$, and the ones calculated with a mode solver and the refractive index profile obtained from Eq. (\ref{diffeq}), $n_{eff}^{calc}$. The wavelength dependence of $\delta$ was determined by interpolating the data acquired at the wavelengths 1550 nm, 780 nm, 635 nm and 532 nm  with the one pole Sellmeier equation $\delta(\lambda)=\sqrt{A+\frac{B}{\lambda^2-C^2}}$.
\begin{table}[b]
\centering\caption{Root-mean-square error $\Delta n_{eff}=\sqrt{\sum_{m=1}^4\left(n_{eff,m}^{calc}-n_{eff,m}^{meas}\right)^2/4}$ in the calculation of the $n_{eff}$ for waveguides shown in Fig. \ref{fig_Ann}(b) and (c). All data refers to the wavelength $\lambda$=635 nm.}
\begin{tabular}{lp{1.0in}|lp{1.0in}}
%\hline
\multicolumn{2}{c|}{X-cut}    &  \multicolumn{2}{|c}{Z-cut}\\ \hline
$t_a$ (h)&  $\Delta n_{eff}\times 10^{-4}$ &   $t_a$ (h)  &  $\Delta n_{eff}\times 10^{-4}$    \\ \hline
36 & 0.6 & 25 & 1.4 \\
44 & 1.8 & 34 & 1.0 \\
59 & 1.2 & 48 & 1.6 \\
\end{tabular}
\label{table_ann}
\end{table}
Table \ref{table_diff} shows the parameters of the diffusion and the coefficients of the Sellmeier equations obtained from the minimization of the rms error $\Delta n_{eff}$.

The Sellmeier curves are shown in Fig. \ref{fig_Ann}(a), while Fig. \ref{fig_Ann}(b) and (c) show the evolution of the refractive index profile during annealing for an X-cut with a PE depth $d_e$ = 0.822 $\mu$m and a Z-cut waveguide with $d_e$ = 0.798 $\mu$m. Table \ref{table_ann} shows the values of the rms error $\Delta n_{eff}$ of the model for these two waveguides characterised by the first four guided modes at $\lambda$ =635 nm.
\subsection{Reverse Proton exchange}
\begin{figure}[t!]
\centering
\includegraphics[width = 0.8\textwidth] {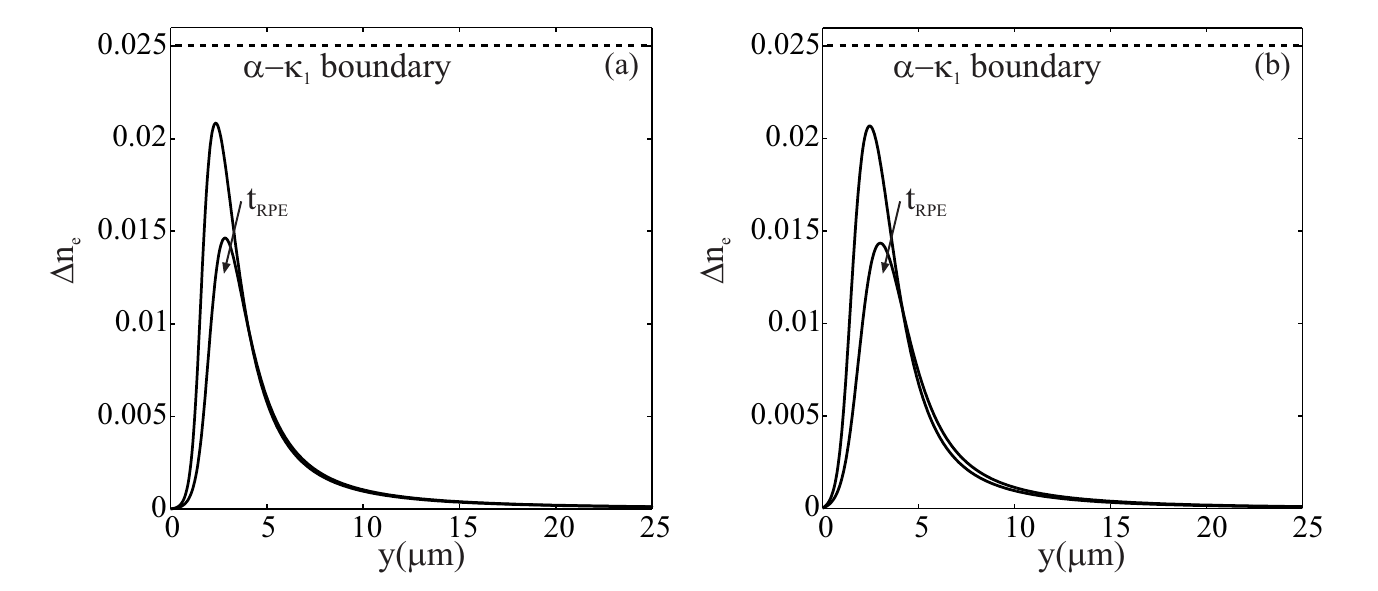}
\caption{(a) Simulated evolution of the refractive index change during RPE for an X-cut planar waveguide calculated form Eq. (\ref{diffeq}). The waveguide had a PE depth $d_e$ = 1.653 $\mu$m and was annealed for $t_a$ = 24 h and reverse proton exchanged for $t_{RPE}$ = 8 h and 11.5 h. (b) Same as (a) but for a Z-cut sample with $d_e$ = 1.926 $\mu$m, $t_a$ = 17 h and $t_{RPE}$ = 10.8 h and 15.6 h.}
\label{fig_RPE}
\end{figure}
The asymmetry in the refractive index profile of APE waveguides (see Fig.\ref{fig_Ann}(b),(c)) generates an asymmetry in the intensity profile of the guided modes that reduces the coupling efficiency with optical fibre and the mode overlap in frequency conversion processes. Furthermore the modes of APE waveguides overlap with the PE dead layer which has suppressed nonlinearity. This problem is overcome with RPE which buries the waveguides through the back substitution of $\mbox{Li}^+$ for $\mbox{H}^+$ at the surface of the crystal.

Reverse Proton Exchange is performed after annealing by dipping the sample in an eutectic melt of $\mbox{LiNO}_3:\mbox{KNO}_3:\mbox{NaNO}_3$ (mole percent ratio of $37.5:44.5:18.0$) at a temperature of $328 \ ^{\circ} \mbox{C}$ \cite{JackelRPE}. During this process the H$^+$ near the LN surface are removed while the other protons are annealed deeper into the substrate.
\begin{table}[b!]
\centering\caption{Root-mean-square error $\Delta n_{eff}$ in the calculation of the $n_{eff}$ for waveguides shown in Fig. \ref{fig_RPE}(a) and (b). All data refers to the wavelength $\lambda$=635 nm.}
\begin{tabular}{lp{1.0in}|lp{1.0in}}
%\hline
\multicolumn{2}{c|}{X-cut}    &  \multicolumn{2}{|c}{Z-cut}\\ \hline
$t_{RPE}$ (h)&  $\Delta n_{eff}\times 10^{-4}$ &   $t_{RPE}$ (h)  &  $\Delta n_{eff}\times 10^{-4}$  \\ \hline
8    & 2.8 & 10.8 & 2.6 \\
11.5 & 1.9 & 15.6 & 1.7
\end{tabular}
\label{table_RPE}
\end{table}
The process is modelled using Eq. (\ref{diffeq}) with changed boundary condition $C=0$ at $y=0$. In this way we model the eutectic melt as a perfectly absorbing layer placed at the $\mbox{LiNbO}_3$ top surface. The parameter $\gamma_{X/Z}$, that controls the value of the diffusion coefficient for $C \to 0$, plays a central role in determining the rate of the RPE process. In fact this rate is mainly affected by the diffusion properties of the protons at the top surface of the crystal, that have concentration values approaching zero. For this reason the values of $\gamma_{X/Z}$ reported in Tab. \ref{table_diff} are obtained through minimization of the RPE data keeping fixed the value of the other parameters obtained by the APE characterization.

Figure \ref{fig_RPE} shows the evolution of the refractive index profile during RPE for X-cut and Z-cut substrates and the rms in the calculation of the $n_{eff}$ for these profiles is given in Tab. \ref{table_RPE}. Reverse exchange moves the peak of the profile below the LN surface causing an improvement in the symmetry of the mode, a reduction in the surface scattering component of the propagation losses and a pulling of the mode away from dead-layer on the surface that is created after PE. %Figure \ref{fig_RPE}a represent a waveguide with a PE depth $d_e$ = 1.653 $\mu$m, annealed for $t_a$ = 24 h and reverse proton exchanged for $t_{RPE}$ = 8 h and 11.5 h. For Fig. \ref{fig_RPE}b e have $d_e$ = 1.926 $\mu$m, $t_a$ = 17 h and $t_{RPE}$ = 10.8 h and 15.6 h.

\section{Design and fabrication of channel waveguides}
\begin{figure}[b]
\includegraphics[width = \textwidth] {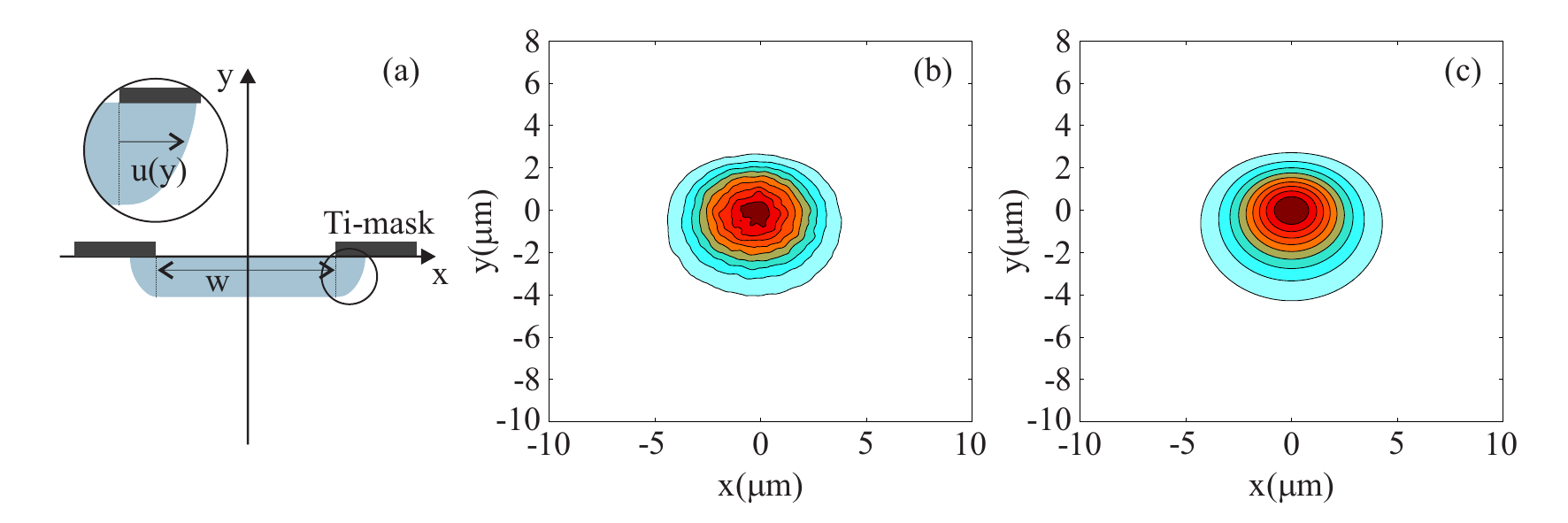}
\caption{(a) Step-like refractive index profile used as initial condition for annealing. The inset shows the undercut diffusion of H$^+$. (b) Measured intensity profile of the guided mode at 1550 nm. (c) Calculated mode from the refractive index profile obtained by solving Eq. (\ref{diffeq2D}). The waveguide had a channel width of $w$=8 $\mu$m, $d_{e}$=1.75 $\mu$m, $t_{a}$=26 h and $t_{RPE}$=14.5 h.}
\label{fig_2D}
\end{figure}
We applied this model for the design and fabrication of straight channel waveguide and directional couplers on a Z-cut substrate. The design constraints were: good coupling with optical fiber, waveguide in the $\alpha$-phase, single mode operation at 1550 nm and low propagation losses.

The fabrication parameters were determined by solving the diffusion equation
\begin{equation}
\frac{\partial C}{\partial t}= \frac{\partial}{\partial y} \left( D_{a,Z} \left( C \right) \frac{\partial C}{\partial y} \right)+\frac{\partial}{\partial x} \left( D_{a,X} \left( C \right) \frac{\partial C}{\partial x} \right),
\label{diffeq2D}
\end{equation}
with the diffusion coefficients obtained in section 2 for annealing and RPE and maximizing our fabrication constraints. The initial condition for the annealing diffusion is the step-like refractive index profile shown in Fig. \ref{fig_2D}(a) where the shaded region has a proton concentration $C=1$. The waveguides were fabricated by patterning a titanium mask on a LN wafer by standard photolithography with a channel width of the waveguides, $w$=8 $\mu$m. The undercut diffusion of proton during PE is modelled by the empirical formula  $u_{under}(y)=u_{max}\sqrt{1-(y/de)^2}$ with $u_{max}=0.2\sqrt{(D_X/D_Z)}d_e$ (see inset in Fig. \ref{fig_2D}(a)).

After our analysis we fabricated the devices using a proton exchange depth $d_{e}$=1.75 $\mu$m, annealing time $t_{a}$=26 h and reverse time $t_{RPE}$=14.5 h. Figure \ref{fig_2D}(b) shows the intensity profile of the waveguide output mode at 1550 nm while Fig \ref{fig_2D}(c) is the mode profile predicted by our model. The overlap between measured and calculated mode is 96\% and overlap between the measured fiber and waveguide modes is 90\% which is in accordance with the 90.5\% prediction from our model. The insertion losses for a 2.8 cm long waveguide were 1.39 dB which account for $0.57\pm0.02$ dB per facet of coupling and Fresnel losses and $0.089\pm0.006$ dB/cm propagation losses.

\begin{figure}[t]
\includegraphics[width = \textwidth] {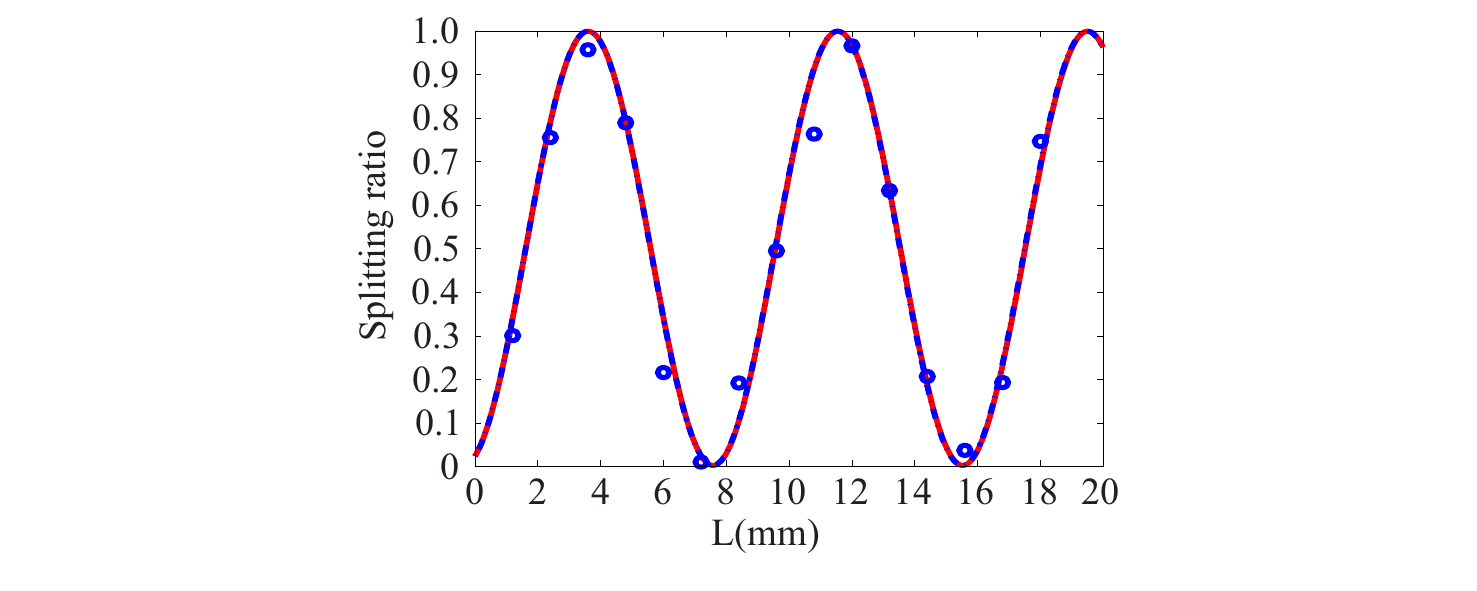}
\caption{Splitting ratio of the directional coupler. $\circ$ are the experimental measurements and the blue-dotted line is their fitting. The red solid line is the function calculated using our model.}
\label{fig_coupling}
\end{figure}

A set of 15 directional couplers with coupling lengths ranging from 1.2 mm to 18 mm were fabricated with the same parameters as the straight waveguide and a separation of 13.5 $\mu$m between the centres of the waveguide in the coupling region. The anisotropy of the diffusion is critical for the modeling of this devices because the lateral diffusion of H$^+$ plays a crucial role in the effective coupling between the waveguides. Figure \ref{fig_coupling} shows the  measured splitting ratios as a function of the coupling length: the data is plotted along with the a sine square fitting function, which gives a coupling length $L_c$ = 3.973$\pm$0.09 mm compared to the estimated value of $L_{c,est}$ = 3.976 mm also shown in Fig. \ref{fig_coupling}. The comparison between measured and simulated splitting ratio has a rms error of 0.06.

\section{Conclusions}
In conclusion we have derived a simple and reliable model of the anisotropic diffusion for the fabrication of APE and RPE waveguide in LN. The model was tested with the design and fabrication of straight waveguides and directional couplers with good agreement between measured and calculated quantities. This model will provide a useful tool for the design and optimization of complex integrated optical devices in LN.
\section*{Acknowledgments}
This work was supported by the Australian Research Council (ARC) under the Grants DP140100808. ML acknowledges the support of the ARC-Decra DE130100304. This work was performed in part at the Griffith node of the Australian National Fabrication Facility. A company established under the National Collaborative Research Infrastructure Strategy to provide nano and microfabrication facilities for Australia’s researchers.
\end{document}